\documentclass[%
 reprint,
 amsmath,amssymb,amsfonts,
 aps,
]{revtex4-2}

\usepackage[dvipdfmx]{graphicx}
\usepackage{dcolumn}
\usepackage{bm}
\usepackage[dvipsnames]{xcolor}

\def\mat#1{#1}

\def\ket#1{\mbox{\boldmath $#1$}}
\def\sket#1{\mbox{\boldmath $\scriptstyle #1$}}

\newcommand{\ER}{Erd\H{o}s--R\'{e}nyi}
\newcommand{\HS}{Hubbard--Stratonovich}
\newcommand{\CH}{Chebyshev--Hermite}
\newcommand{\BC}{Bender--Canfield}

\begin{document}


\title{Entropy of microcanonical finite-graph ensembles}


\author{Tatsuro Kawamoto}
\affiliation{Artificial Intelligence Research Center, \\
  National Institute of Advanced Industrial Science and Technology, 
  Tokyo, Japan }



\begin{abstract}
The entropy of random graph ensembles has gained widespread attention in the field of graph theory and network science. 
We consider microcanonical ensembles of simple graphs with prescribed degree sequences. 
We demonstrate that the mean-field approximations of the generating function using the Chebyshev--Hermite polynomials provide estimates for the entropy of finite-graph ensembles. 
Our estimate reproduces the Bender--Canfield formula in the limit of large graphs. 
\end{abstract}


\maketitle

\section{Introduction}
The entropy of physical systems is a fundamental quantity in statistical physics. 
For microcanonical ensembles, the entropy is represented as the logarithm of the total number of realizations or states in the system (Boltzmann's principle \cite{TodaStatMech}). 
When the system is represented as a graph, a microcanonical ensemble is a set of graphs wherein every graph satisfies specific macroscopic constraints. 
Calculating the entropy amounts to calculating the total number of graphs allowed in the ensemble, denoted as $\mathcal{N}_{G}$ in this study. 
By contrast, a canonical ensemble is a set of graphs wherein the macroscopic constraints are satisfied as an expectation. 
Throughout this study, we only consider microcanonical graph ensembles that include simple (no multi-edges or self-loops) undirected graphs. 

Initially, the entropy of the graph ensembles received attention in graph theory, particularly in the 1980s \cite{Bollobas2010handbook,Wormald1999}. 
The exact number of graphs in an ensemble cannot be expressed in a compact form in many cases, and several bounds and estimates have been proposed in the literature \cite{Wormald1999}. 
One of the early results is an asymptotic formula referred to as the {\BC} formula \cite{BenderCanfield1978}, which is introduced later in this study. 
Although the original study considered only graphs with degrees of $O(1)$, finer conditions for the formula were considered in later studies \cite{Bollobas1980,McKay1985,MckayWormald1991,MckayWormald1997,Frieze2016}. 
Statistical physicists also studied the entropy of graph ensembles. 
Bianconi et al. \cite{Bianconi2007,Bianconi2009,Bianconi2008,AnandBianconi2009} derived entropies for graph ensembles with constraints that are often considered in the field of network science, such as degree correlations, community structures, and spatial embedding structures. 
In Ref.~\cite{Peixoto2012}, Peixoto used a combinatorial approach to evaluate the entropy of stochastic block models. 
See Refs.~\cite{RobertsCoolen2014,RobertsCoolen2014_JPhysA,squartini2017maximum,Lopez2018,Lopez2021,Paton2022,Bianconi2022} for other recent related works in the physics community. 

The entropy of microcanonical ensembles provides the probability of instances in randomized graphs. 
Thus, entropy and related quantities (e.g., moments) can be useful for the statistical evaluation of random graph models and are also related to the graph sampling problem \cite{DelGenio2010,Squartini2015}. 
Moreover, the entropy of stochastic block models \cite{Peixoto2012,Peixoto2017,Peixoto2017tutorial,PeixotoPRX2022,PeixotoPRE2022} can be utilized for the statistical inference of module structures. 
We note that entropy is a primary quantity in statistical physics, and we estimate the number of graphs because entropy is obtained as the logarithm of the number of graphs. 

This study focuses on ensembles of finite-size graphs, particularly those with $\mathcal{N}_{G} < 10^{6}$, which often correspond to graphs with less than ten vertices. 
We consider the mean-field estimates of $\mathcal{N}_{G}$, while maintaining a connection with its combinatorial estimations. 
After deriving a formula for finite graphs, a large size limit is considered to verify the reproduction of the asymptotic formula. 

The rest of the study is structured as follows. 
Section~\ref{sec:NumGraphs} introduces a generating function approach to the calculation of the number of graphs by considering a simple example. 
Whereas generating functions are common in the literature, their mean-field estimates for finite-graph ensembles have not been investigated. 
Section~\ref{sec:MeanField} considers the mean-field estimates for a nontrivial graph ensemble. 
The second-order (Sec.~\ref{sec:NaiveApprox}) and corrected second-order (Sec.~\ref{sec:SecondOrderExpansion}) approximations provide combinatorially interpretable estimates. 
In contrast, the estimate with the fourth-order approximation (Sec.~\ref{sec:FourthOrderExpansion}) is novel, combinatorially nontrivial, and sufficient to reproduce the {\BC} formula. 
Section \ref{sec:Experiments} evaluates the accuracy of each approximation for small graph ensembles through numerical experiments. 
Finally,  Sec.~\ref{sec:Discussion} is devoted to the discussion of the obtained results.

\section{Number of graphs}\label{sec:NumGraphs}
For a graph $G = (V, E)$ with vertices $V$ and edges $E$, we denote $|V|=N$ and $|E|=M$. 
We denote the adjacency matrix of a graph as $\mat{A}$; $A_{ij}=1$ when vertex pair $i$ and $j$ is connected and $A_{ij}=0$ otherwise. 
Because we consider undirected simple graphs, the adjacency matrix is a binary symmetric matrix. 
We assume that each vertex is distinguishable (labeled graph), and hence, graph isomorphisms are counted as different graph instances. 

Let $\{ A_{ij} \in \{0,1\} \}$ be the set of $2^{\binom{N}{2}}$ configurations in the upper-right part of the adjacency matrix elements. 
The number of graphs $\mathcal{N}_{G}$ can be formally counted as the sum over $\{ A_{ij} \in \{0,1\} \}$ with constraints. 
For example, when the total number of edges is constrained to $M$, 
\begin{align}
\mathcal{N}_{G} = \sum_{\{ A_{ij} \in \{0,1\} \}} \delta\biggl( \sum_{i<j}A_{ij}, M \biggr), \label{ERrandomgraph}
\end{align}
where $\sum_{i<j}A_{ij}$ is the sum of all upper-right elements in $\mat{A}$ and $\delta(x,y)$ represents the Kronecker delta, which is one when $x=y$ and zero otherwise. 
In the {\ER} random graph model \cite{ErdosRenyi1959,Bollobas2001Book}, each instance in the $\mathcal{N}_{G}$ graphs is realized with equal probability. 

The Kronecker delta has the following integral representation: 
\begin{align}
\delta\left( x, y \right) &= \oint \frac{dz}{2\pi i} \, z^{x-y-1}. \label{KroneckerDelta}
\end{align}
where $x$ and $y$ are discrete variables. 
In Eq.~(\ref{KroneckerDelta}), the integral is along an arbitrary loop around the origin in the complex plane; the integral is one only when the exponent of $z$ is equal to one, and zero otherwise, according to Cauchy's integral theorem. 
Using this trick and the residue theorem, we obtain a combinatorial expression for $\mathcal{N}_{G}$ in Eq.~(\ref{ERrandomgraph}): 
\begin{align}
\mathcal{N}_{G} 
&= \oint \frac{dz}{2\pi i} \frac{1}{z^{1+M}} \prod_{i<j} \biggl( \sum_{A_{ij} \in \{0,1\}} z^{A_{ij}} \biggr) 
= \oint \frac{dz}{2\pi i} \frac{(1+z)^{\binom{N}{2}}}{z^{1+M}} \notag\\
&= \frac{1}{M!} \left. \frac{d^{M}}{z^{M}} (1+z)^{\binom{N}{2}} \right|_{z=0} 
= \binom{\binom{N}{2}}{M}. 
\end{align}
This is a simple example of counting the number of graphs. 
We can clearly interpret this solution through the combinatorial expression; that is, $\mathcal{N}_{G}$ is simply the number of ways to select $M$ nonzero elements from the $\binom{N}{2}$ upper-right cells of the adjacency matrix elements. 
However, exact combinatorial expressions cannot always be obtained in compact and interpretable forms. 
We must rely on the approximations in such cases, and we consider such cases for the remainder of this study.

\section{Mean-field solutions for graphs with prescribed degree sequence}\label{sec:MeanField}
We evaluate the number of graphs in the ensemble where the degree sequence is constrained. 
The degree $d_{j}$ of vertex $j$ is expressed as $\sum_{i}A_{ij}$. 
We denote $\{d_{1}, \dots, d_{N} \} =: \ket{d}$ for the degree sequence. 
The total number of graphs $\mathcal{N}_{G}(\ket{d})$ is 
\begin{align}
\mathcal{N}_{G}(\ket{d}) 
&= \sum_{\{ A_{ij} \in \{0,1\} \}} \prod_{j=1}^{N} \delta\biggl( d_{j}, \sum_{i=1}^{N} A_{ij} \biggr) \notag\\
&= \oint \prod_{j=1}^{N}\left(\frac{dz_{j}}{2\pi i} \frac{1}{z_{j}^{1+d_{j}}}\right) \mathrm{e}^{F(\ket{z})} \notag\\
&= \left. \frac{1}{\prod_{j=1}^{N} d_{j}!} \frac{\partial^{d_{1}} \cdots \partial^{d_{N}}}{\partial z_{1}^{d_{1}} \cdots \partial z_{N}^{d_{N}}} \mathrm{e}^{F(\ket{z})} \right|_{z_{1}=0, \dots, z_{N}=0} 
\label{NumGraphs_GeneratingFunction}
\end{align}
where
\begin{align}
F(\ket{z})
= \sum_{i < j} \log\left( \sum_{A_{ij} \in \{0,1\}} (z_{i}z_{j})^{A_{ij}} \right) 
= \sum_{i < j} \log\left( 1 + z_{i}z_{j} \right).
\end{align}
Here, $\ket{z}$ represents the set of auxiliary variables $\{z_{1}, \dots, z_{N}\}$. 
The final expression in Eq.~(\ref{NumGraphs_GeneratingFunction}) indicates that $\mathrm{e}^{F(\ket{z})}$ is a generating function of $\mathcal{N}_{G}(\ket{d})$. 
The general solution for a higher-order derivative can be provided in the form of the Fa\`{a} di Bruno's formula \cite{johnson2003curious}. Nevertheless, approximations are often required, because this formula involves a complicated sum. 

$\mathcal{N}_{G}(\ket{d})$ is a nontrivial quantity and an asymptotic solution was found in graph theory. 
The so-called {\BC} formula estimates that 
\begin{align}
\mathcal{N}^{\mathrm{BC}}_{G}(\ket{d}) = \frac{(2M-1)!!}{\prod_{j=1}^{N} d_{j}!} \, \mathrm{e}^{-\lambda-\lambda^{2}}, \label{BCformula}
\end{align}
where 
\begin{align}
\lambda = \frac{1}{2M} \sum_{j=1}^{N} \binom{d_{j}}{2}. \label{def-lambda}
\end{align}
This solution becomes accurate when $N$ is sufficiently large as long as $\max_{j} d_{j}$ is sufficiently small (see Ref.~\cite{Frieze2016} for a more precise argument on the scaling of degrees).
Here, the double factorial is defined as $n!! \equiv \prod_{k=0}^{\lceil n/2 \rceil-1} (n-2k) = n (n-2) (n-4) \cdots$ and $\lceil \cdot \rceil$ is the ceil function. 
The standard derivation of Eq.~(\ref{BCformula}) is based on combinatorial and probabilistic arguments (see Refs.~\cite{Wormald1999,Noy2015} for details). 
In contrast, we consider mean-field approximations of $F(\ket{z})$ and examine the resulting estimate of the number of graphs. 

\subsection{Second-order approximation}\label{sec:NaiveApprox}
First, we consider the following second-order approximation for $F(\ket{z})$: 
\begin{align}
F_{\mathrm{2nd}}(\ket{z})
= \frac{1}{2}\biggl( \sum_{j=1}^{N} z_{j} \biggr)^{2}. \label{MeanFieldFreeEnergy}
\end{align}
This is attributed to the lowest-order expansion $\sum_{i<j}\log\left( 1 + z_{i}z_{j} \right) \simeq \sum_{i<j} z_{i}z_{j}$. 
Note that $\sum_{j} z_{j}^{2}/2$ must be subtracted from Eq.~(\ref{MeanFieldFreeEnergy}) to be more precise. 
Although this correction is considered in the next section, it is neglected assuming that its contribution is not dominant. 

Using the {\HS} transform, 
\begin{align}
\mathrm{e}^{F_{\mathrm{2nd}}(\ket{z})} 
= \mathrm{e}^{\frac{1}{2}(\sum_{j} z_{j})^{2}} 
= \int^{+\infty}_{-\infty} \frac{dQ}{\sqrt{2\pi}} \mathrm{e}^{-\frac{Q^{2}}{2} + Q \sum_{j} z_{j}}, 
\end{align}
the number of graphs $\mathcal{N}^{\mathrm{2nd}}_{G}(\ket{d})$ under this approximation can be evaluated as 
\begin{align}
\mathcal{N}^{\mathrm{2nd}}_{G}(\ket{d}) 
&= \int \frac{dQ}{\sqrt{2\pi}} \mathrm{e}^{-\frac{Q^{2}}{2}} 
\prod_{j=1}^{N} \left(\oint \frac{dz_{j}}{2\pi i} \frac{\mathrm{e}^{Qz_{j}}}{z_{j}^{1+d_{j}}}\right) \notag\\
&= \int \frac{dQ}{\sqrt{2\pi}} \mathrm{e}^{-\frac{Q^{2}}{2}} 
\prod_{j=1}^{N} \left( \left. \frac{1}{d_{j}!} \frac{\partial^{d_{j}} \mathrm{e}^{Qz_{j}}}{\partial z_{j}^{d_{j}}} \right|_{z_{j}=0}\right) \notag\\
&= \frac{1}{\prod_{j}d_{j}!} \int \frac{dQ}{\sqrt{2\pi}} \, Q^{\sum_{j} d_{j}} \mathrm{e}^{-\frac{Q^{2}}{2}} 
= \frac{(2M -1 )!!}{\prod_{j}d_{j}!}. \label{NumGraphs_MF1}
\end{align}
Note that $\sum_{j}d_{j} = 2M$. 
In the last equality, an identity of the Gaussian integral, $\int^{+\infty}_{-\infty} dx \, x^{n} \mathrm{e}^{-\frac{x^{2}}{2}} = (n-1)!!$ for an even integer $n$ is used. 

\begin{figure*}[t!]
    \includegraphics[width=1.75\columnwidth]{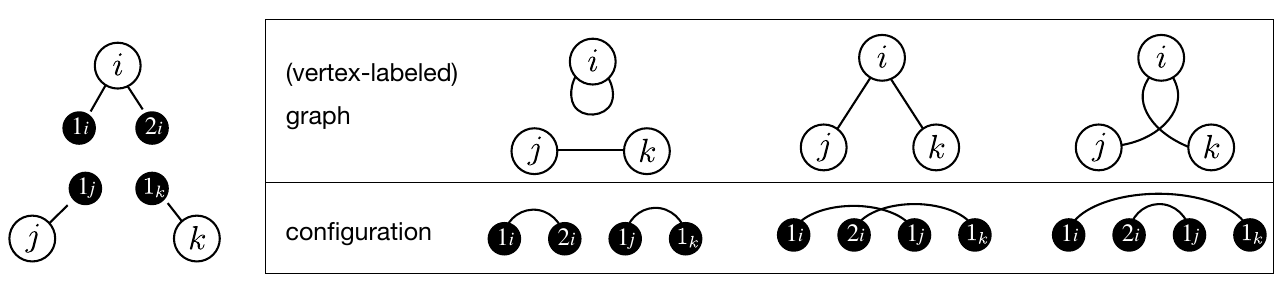}
    \caption{
    Possible graphs and the corresponding stub-labeled graphs (configurations) for a degree sequence. 
    The white circles represent three vertices $i$, $j$, and $k$. 
    The black circles represent stubs and the vertex that each stub is incident on is indicated as a subscript of the index label. 
    }
    \label{fig:configurations_3vertices}
\end{figure*}

A combinatorial explanation for Eq.~(\ref{NumGraphs_MF1}) can be obtained by considering the stub-labeled graphs (also called the configurations \cite{Bollobas1980,Peixoto2012}). 
A stub is a ``half-edge'' that is incident on a vertex. 
Every stub is distinguishable in a stub-labeled graph. 
For a given degree sequence, the number of stub-labeled graphs can be counted by pairing all $2M$ stubs. 
This pairing is known as the perfect matching, and the number of perfect matching is $(2M-1)!!$. 
A specific example is shown in Fig.~\ref{fig:configurations_3vertices}. 
However, this is an overestimation because the stubs are indistinguishable in the considered graphs. 
That is, different stub-labeled graphs are obtained by permuting the stubs incident on the same vertex. 
The factor $\prod_{j}d_{j}!$ corrects these degrees of freedom. 

The solution is not precise even after the correction by $\prod_{j}d_{j}!$. 
When the perfect matching is considered in the stub pairing, the vertex on which each stub is incident is ignored. 
The pairings represent simple graphs as well as graphs with multi-edges and self-loops. 
Therefore, finer estimates are required.

\subsection{Corrected second-order expansion}\label{sec:SecondOrderExpansion}
Next, we consider a slightly more accurate expansion of $F_{\mathrm{2nd}}(\ket{z})$ in which the contribution from the diagonal elements in Eq.~(\ref{MeanFieldFreeEnergy}) is corrected: 
\begin{align}
F_{\mathrm{2nd+}}(\ket{z}) = \frac{1}{2}\biggl( \sum_{j=1}^{N} z_{j} \biggr)^{2} - \frac{1}{2} \sum_{j=1}^{N} z_{j}^{2}. \label{HermiteFreeEnergy}
\end{align}

Applying the {\HS} transform again, we obtain 
\begin{align}
\mathcal{N}^{\mathrm{2nd+}}_{G}(\ket{d}) 
&= \int \frac{dQ}{\sqrt{2\pi}} \mathrm{e}^{-Q^{2}/2} \prod_{j=1}^{N}\left(\oint \frac{dz_{j}}{2\pi i} \frac{\mathrm{e}^{-z_{j}^{2}/2 + Q z_{j}} }{z_{j}^{1+d_{j}}}\right) \notag\\
&= \frac{1}{\prod_{j=1}^{N} d_{j}!} \int \frac{dQ \, \mathrm{e}^{-\frac{Q^{2}}{2}}}{\sqrt{2\pi}} \prod_{j=1}^{N} He_{d_{j}}(Q) \notag\\
&=: \frac{\mathcal{J}(d_{1}, \dots, d_{N})}{\prod_{j=1}^{N} d_{j}!}. \label{NumGraphs_MF2_1}
\end{align}
Here, $He_{n}(x)$ is the {\CH} polynomial: 
\begin{align}
He_{n}(x) 
&= n! \oint \frac{dz}{2\pi i} \frac{\mathrm{e}^{-\frac{1}{2}z^{2} + x z} }{z^{1+n}} \notag\\
&= n! \sum_{m=0}^{\lfloor n/2 \rfloor} \frac{(-1)^{m}}{m! (n-2m)!} \frac{x^{n-2m}}{2^{m}}, \label{CHPolynomial}
\end{align}
where $\lfloor \cdot \rfloor$ is the floor function. 
$\mathcal{J}(d_{1}, \dots, d_{N})$ represents the Gaussian integral of the product of $N$ {\CH} polynomials. 
$\mathcal{J}(d_{1}, \dots, d_{N})$ is equivalent to the number of perfect matchings in a complete $N$-partite graph (a graph with $N$ independent vertex sets) \cite{andrews1999special}. 

Equation (\ref{NumGraphs_MF1}) considered the pairing of stubs without distinguishing the vertex on which each stub is incident. 
In contrast, Eq.~(\ref{NumGraphs_MF2_1}) distinguishes the vertices to prohibit the self-loops in its pairing. 
However, this counting is still not precise, because multi-edges are not prohibited. 

Let us express Eq.~(\ref{NumGraphs_MF2_1}) using the second equality in Eq.~(\ref{CHPolynomial}): 
\begin{align}
&\mathcal{N}^{\mathrm{2nd+}}_{G}(\ket{d}) \notag\\
&= \sum_{k_{1}=0}^{\lfloor d_{1}/2 \rfloor} \cdots \sum_{k_{N}=0}^{\lfloor d_{N}/2 \rfloor} 
\frac{ \int_{-\infty}^{+\infty} \frac{dQ \, \mathrm{e}^{-\frac{Q^{2}}{2}}}{\sqrt{2\pi}} Q^{2 (\sum_{j} d_{j}/2 - \sum_{j} k_{j})} }{\prod_{j} k_{j}! (d_{j} - 2k_{j})! (-2)^{k_{j}}} \notag\\
&= \sum_{k_{1}=0}^{\lfloor d_{1}/2 \rfloor} \cdots \sum_{k_{N}=0}^{\lfloor d_{N}/2 \rfloor} 
\frac{\left( 2(M-\sum_{j}k_{j})-1 \right)!!}{\prod_{j} k_{j}! (d_{j} - 2k_{j})! (-2)^{k_{j}}} \notag\\
&= \frac{1}{\prod_{j=1}^{N} d_{j}!} \, 
\sum_{K=0}^{M} (-1)^{K} (2(M-K)-1)!! \notag\\
&\quad\quad\quad\quad\times \sum_{\sket{k}(\sket{d}, K)} 
\prod_{j=1}^{N} \binom{d_{j}}{2k_{j}} (2k_{j}-1)!!,
\label{NumGraphs_MF2_3-1} 
\end{align}
where we used the Gaussian identity of the double factorial again and introduced 
\begin{align}
\sum_{\sket{k}(\sket{d}, K)} := \sum_{k_{1}=0}^{\lfloor d_{1}/2 \rfloor} \cdots \sum_{k_{N}=0}^{\lfloor d_{N}/2 \rfloor} \delta\biggl(\sum_{j=1}^{N} k_{j}, K \biggr) \label{ConstrainedSum}
\end{align}
to simplify the notation. Note that $0!! = (-1)!! = 1$. 
$\ket{k}(\ket{d}, K) = \{k_{1}, \dots, k_{N}\}$ represents a sequence of non-negative integers that is constrained locally by $\lfloor d_{j}/2 \rfloor$ for each vertex and globally by $\sum_{j=1}^{N} k_{j} = K$. 
Equation (\ref{NumGraphs_MF2_3-1}) is an expansion with respect to the least number of self-loops. 
When $K>0$ is ignored, it coincides with $\mathcal{N}^{\mathrm{2nd}}_{G}(\ket{d})$. 
For a given $\ket{k}(\ket{d}, K)$, 
\begin{align}
( 2(M-K)-1 )!! 
\prod_{j=1}^{N} \binom{d_{j}}{2k_{j}} (2k_{j}-1)!! \label{ConfigurationsWithSelfLoops}
\end{align}
is the number of stub-labeled graphs in which each vertex has at least $k_{j}$ self-loops.
Note that the self-loops are still included in Eq.~(\ref{ConfigurationsWithSelfLoops}) within the pairing of $2(M-K)$ stubs. 
In this term, the number of stub-labeled graphs with at least $K+1$ self-loops is overcounted. 
The $(K+1)$th term corrects this overcounting, although the term again includes some overcounting, and so on. 
In the end, the entire sum counts the pairing of stubs without self-loops. 

To see how Eq.~(\ref{NumGraphs_MF2_3-1}) is related to Eq.~(\ref{BCformula}), we assume that $N$ and $M$ are sufficiently large and approximate Eq.~(\ref{NumGraphs_MF2_3-1}) as 
\begin{align}
&\mathcal{N}^{\mathrm{2nd+}}_{G}(\ket{d}) \notag\\
&= \frac{(2M-1)!!}{\prod_{j=1}^{N} d_{j}!} \, 
\Biggl( 
1 - \frac{\sum_{j=1}^{N} \binom{d_{j}}{2}}{2M-1} \notag\\
&+ \frac{ \frac{1}{2!}\left(\sum_{j=1}^{N} \binom{d_{j}}{2} \right)^{2} + \sum_{j=1}^{N} \left( 3\binom{d_{j}}{4} - \frac{1}{2!}\binom{d_{j}}{2}^{2} \right)}{(2M-1)(2M-3)} 
- \cdots
 \Biggr) \notag\\
&\approx \frac{(2M-1)!!}{\prod_{j=1}^{N} d_{j}!} \, 
\left( 
\sum_{K=0}^{\infty} \frac{1}{K!} \left(-\frac{\sum_{j=1}^{N} \binom{d_{j}}{2}}{2M}\right)^{K}
 \right) \notag\\
&= \frac{(2M-1)!!}{\prod_{j=1}^{N} d_{j}!} \, \mathrm{e}^{-\lambda}, \label{NumGraphs_MF2_3-2} 
\end{align}
where $\lambda$ is defined in Eq.~(\ref{def-lambda}). 
In the second equality, we neglected the contributions of the sums with $o(N^{K})$ elements in the numerator and approximated that $(2M-1)\cdots(2(M-K)+1) \approx (2M)^{K}$ in the denominator. 
Compared with Eq.~(\ref{NumGraphs_MF1}), Eq.~(\ref{NumGraphs_MF2_3-2}) includes $\mathrm{e}^{-\lambda}$ as a correction factor, which partially reproduces Eq.~(\ref{BCformula}). 
In the standard derivation of the {\BC} formula, $\mathrm{e}^{-\lambda}$ is interpreted as a factor based on the probability that a graph includes self-loops in a uniformly random graph-generation process. 
The present derivation also indicates that this correction can be attributed to the exclusion of self-loops. 
Next, we reproduce the remaining correction factor $\mathrm{e}^{-\lambda^{2}}$ in Eq.~(\ref{BCformula}).

\subsection{Fourth-order expansion}\label{sec:FourthOrderExpansion}
Let us consider a fourth-order expansion of $F(\ket{z})$: 
\begin{align}
F_{\mathrm{4th}}(\ket{z})
\simeq \frac{1}{2}\biggl( \sum_{j=1}^{N} z_{j} \biggr)^{2} - \frac{1}{2} \sum_{j=1}^{N} z_{j}^{2} 
- \frac{1}{4}\biggl( \sum_{j=1}^{N} z^{2}_{j} \biggr)^{2}. \label{HermiteFreeEnergy3}
\end{align}
Note that this is not a precise fourth-order expansion with respect to $\{ z_{j} \}$ because the contribution from $\sum_{j} z^{4}_{j}/4$ is not corrected. 
However, as shown below, this is sufficient for reproducing Eq.~(\ref{BCformula}) in the asymptotic regime. 

Once again, the {\HS} transform is used to estimate the number of graphs $\mathcal{N}^{\mathrm{4th}}_{G}(\ket{d})$ as follows: 
\begin{align}
&\mathcal{N}^{\mathrm{4th}}_{G}(\ket{d}) \notag\\
&= \oint \prod_{j=1}^{N}\left(\frac{dz_{j}}{2\pi i} \frac{\mathrm{e}^{-\frac{1}{2} z_{j}^{2}} }{z_{j}^{1+d_{j}}}\right) 
\exp\left( \frac{1}{2}\left( \sum_{j=1}^{N} z_{j} \right)^{2} 
-\frac{1}{4}\left( \sum_{j=1}^{N} z^{2}_{j} \right)^{2} \right) \notag\\
&= \int\frac{dQ_{1} dQ_{2}}{2\pi}  \mathrm{e}^{-\frac{Q^{2}_{1} + Q^{2}_{2}}{2} }
\prod_{j=1}^{N} \left( \oint \frac{dz_{j}}{2\pi i} \frac{ \mathrm{e}^{-\frac{1}{2}(1+i\sqrt{2}Q_{2})z_{j}^{2} + Q_{1}z_{j} } }{z_{j}^{1+d_{j}}} \, 
\right)\notag\\
&= \frac{1}{ \prod_{j} d_{j}! } 
\int\frac{dQ_{1} dQ_{2}}{2\pi}  \mathrm{e}^{-\frac{Q^{2}_{1} + Q^{2}_{2}}{2} } 
\left( 1 + i\sqrt{2}Q_{2} \right)^{M} \notag\\
&\quad\times\prod_{j=1}^{N} He_{d_{j}}\left( \frac{Q_{1}}{\sqrt{1 + i\sqrt{2} Q_{2}}} \right). \label{NumGraphs_MF4_1} 
\end{align}
The expansion of the {\CH} polynomial yields 
\begin{align}
\mathcal{N}^{\mathrm{4th}}_{G}(\ket{d}) 
&= \sum_{k_{1}=0}^{\lfloor d_{1}/2 \rfloor} \cdots \sum_{k_{N}=0}^{\lfloor d_{N}/2 \rfloor} 
\frac{1}{\prod_{j} k_{j}! (d_{j} - 2k_{j})! (-2)^{k_{j}}} \notag\\
&\quad\times\int_{-\infty}^{+\infty} \frac{dQ_{1}}{\sqrt{2\pi}}  \, \mathrm{e}^{-\frac{Q^{2}_{1}}{2}} Q_{1}^{2 (M - \sum_{j} k_{j})} \notag\\
&\quad\times\int_{-\infty}^{+\infty} \frac{dQ_{2}}{\sqrt{2\pi}} \mathrm{e}^{-\frac{1}{2} Q^{2}_{2}} 
\left( 1 + i\sqrt{2}Q_{2} \right)^{\sum_{j} k_{j}}
\notag\\
&= \sum_{k_{1}=0}^{\lfloor d_{1}/2 \rfloor} \cdots \sum_{k_{N}=0}^{\lfloor d_{N}/2 \rfloor} 
\frac{ \left( 2(M-\sum_{j}k_{j})-1 \right)!! }{\prod_{j} k_{j}! (d_{j} - 2k_{j})! (-2)^{k_{j}}} \notag\\
&\quad\times\int \frac{dQ_{2}}{\sqrt{2\pi}} \mathrm{e}^{-\frac{1}{2} Q^{2}_{2}} 
\left( 1 + i\sqrt{2}Q_{2} \right)^{\sum_{j} k_{j}}. \label{NumGraphs_MF4_2}
\end{align}
The former part of the above equation is equivalent to Eq.~(\ref{NumGraphs_MF2_3-1}). 
The integral in the latter part is 
\begin{align}
&\int \frac{dQ_{2}}{\sqrt{2\pi}} \mathrm{e}^{-\frac{1}{2} Q^{2}_{2}} 
( 1 + i\sqrt{2}Q_{2} )^{K} 
= H_{K}\left( 1/2 \right), \label{NumGraphs_MF4_3}
\end{align}
where $K = \sum_{j} k_{j}$ and $H_{n}(x)$ is the (physicist's) Hermite polynomial, which is related to the {\CH} polynomial as $H_{n}(x) = 2^{n/2} He_{n}(\sqrt{2} x)$. 
Analogous to Eq.~(\ref{NumGraphs_MF2_3-1}), Eq.~(\ref{NumGraphs_MF4_2}) is recast as 
\begin{align}
\mathcal{N}^{\mathrm{4th}}_{G}(\ket{d})
&= \frac{1}{\prod_{j=1}^{N} d_{j}!} \, 
\sum_{K=0}^{M} (-1)^{K} H_{K}\left( 1/2 \right) (2(M-K)-1)!! \notag\\
&\quad\quad\quad\quad\times \sum_{\sket{k}(\sket{d}, K)} 
\prod_{j=1}^{N} \binom{d_{j}}{2k_{j}} (2k_{j}-1)!!. \label{NumGraphs_MF4_4}
\end{align}
In the asymptotic regime, by applying the same approximation as in Eq.~(\ref{NumGraphs_MF2_3-2}), we obtain 
\begin{align}
\mathcal{N}^{\mathrm{4th}}_{G}(\ket{d}) 
&\approx \frac{(2M-1)!!}{\prod_{j=1}^{N} d_{j}!} \, 
\sum_{K=0}^{\infty} \frac{H_{K}\left( 1/2 \right)}{K!} \left(-\frac{\sum_{j=1}^{N} \binom{d_{j}}{2}}{2M}\right)^{K} \notag\\
&= \frac{(2M-1)!!}{\prod_{j=1}^{N} d_{j}!} \, \mathrm{e}^{-\lambda-\lambda^{2}}. \label{NumGraphs_MF4_5}
\end{align}
Here, we used the identity 
\begin{align}
\mathrm{e}^{2xt - t^{2}} = \sum_{n=0}^{\infty} \frac{t^{n}}{n!} H_{n}(x). \label{HermiteExpansionOfGaussian}
\end{align}

The fourth-order expansion reproduced Eq.~(\ref{BCformula}). 
In the standard derivation of the {\BC} formula, the correction factor $\mathrm{e}^{-\lambda^{2}}$ is interpreted as a factor based on the probability that a graph includes multi-edges with the multiplicity of two in a uniformly random graph-generation process. 
In contrast, Eq.~(\ref{NumGraphs_MF4_4}) is still an expansion with respect to the least number of self-loops $K$, but with a correction factor $H_{K}\left( 1/2 \right)$. 
Although this correction factor should be related to the exclusion of multi-edges, Eq.~(\ref{NumGraphs_MF4_4}) does not explicitly corrects the contribution of the multi-edges. 
Interestingly, this approximation is sufficient for reproducing Eq.~(\ref{BCformula}) in the asymptotic regime. 
To the best of our knowledge, the series expansion (\ref{NumGraphs_MF4_4}) has not been previously reported, and as demonstrated in the next section, it is often more accurate than the asymptotic form (\ref{BCformula}) when $N$ is small.

\section{Experiments}\label{sec:Experiments}
While the fourth-order expansion yields an estimate $\mathcal{N}^{\mathrm{4th}}_{G}(\ket{d})$ of $\mathcal{N}_{G}(\ket{d})$ for finite graphs, it is not obvious whether the estimate is closer to the exact value than $\mathcal{N}^{\mathrm{BC}}_{G}(\ket{d})$. 
Furthermore, the amount of deviations of $\mathcal{N}^{\mathrm{2nd}}_{G}(\ket{d})$ and $\mathcal{N}^{\mathrm{2nd+}}_{G}(\ket{d})$ from $\mathcal{N}_{G}(\ket{d})$ for small graphs is noteworthy. 
Note that our enumeration does not require the graphs to be connected; unless very dense graph ensembles are considered, some of the graphs comprise disconnected components. 
Thus, the total number of vertices $N$ in a graph is always fixed to the specified number. 
Note also that although the summation (\ref{ConstrainedSum}) is apparently complicated, this operation can be easily coded using the \texttt{itertools.product} function in Python. 

\begin{figure}[t!]
    \includegraphics[width=0.9\columnwidth]{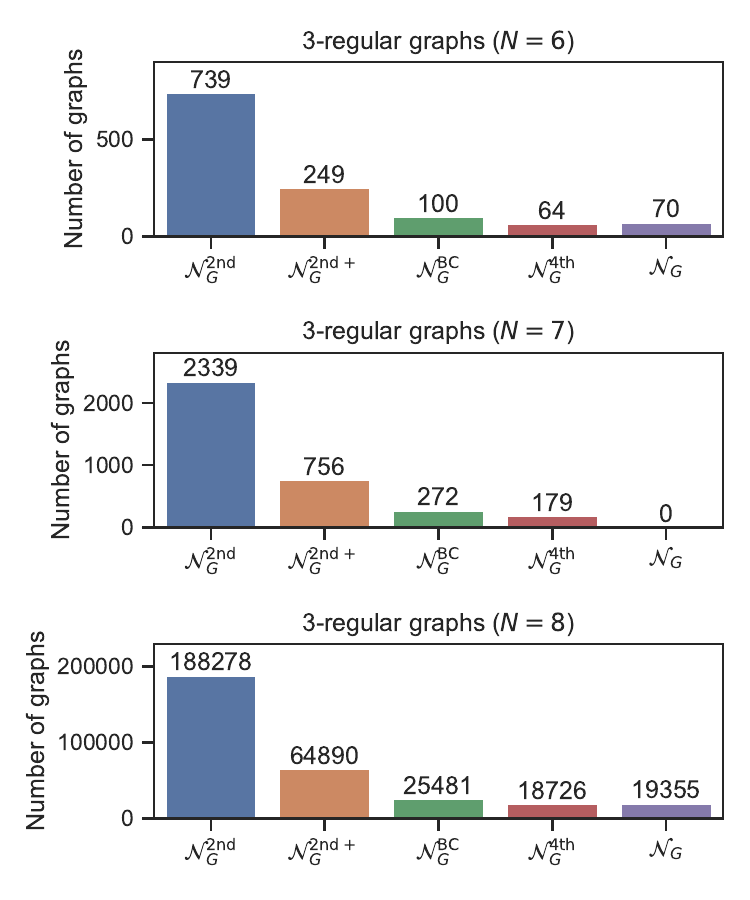}
    \caption{Estimates of the number of graphs in the microcanonical $3$-regular random graph model of different sizes.
    $\mathcal{N}^{\mathrm{2nd}}_{G}$, $\mathcal{N}^{\mathrm{2nd+}}_{G}$, $\mathcal{N}^{\mathrm{BC}}_{G}$, and $\mathcal{N}^{\mathrm{4th}}_{G}$ are respectively the estimates in Eqs.~(\ref{NumGraphs_MF1}), (\ref{NumGraphs_MF2_3-1}), (\ref{BCformula}), and (\ref{NumGraphs_MF4_4}). 
    $\mathcal{N}_{G}$ is the exact number of graphs in the ensemble. 
    }
    \label{fig:3regular}
\end{figure}

Figure \ref{fig:3regular} compares, for a given $N$, the estimates of the number of graphs in the microcanonical ensemble of $3$-regular graphs with the exact value $\mathcal{N}_{G}$, which is counted through the brute-force enumeration. 
Whereas $\mathcal{N}^{\mathrm{2nd}}_{G}(\ket{d})$ and $\mathcal{N}^{\mathrm{2nd+}}_{G}(\ket{d})$ largely overestimate the number of graphs, $\mathcal{N}^{\mathrm{BC}}_{G}(\ket{d})$ is closer to $\mathcal{N}_{G}(\ket{d})$, and $\mathcal{N}^{\mathrm{4th}}_{G}(\ket{d})$ is even closer to $\mathcal{N}_{G}(\ket{d})$ for $N=6,7$, and $8$. 
For $N=7$, $\mathcal{N}_{G}(\ket{d})=0$ because graphs cannot be constructed when the total degree is odd. 
The mean-field estimates do not take this feature into account. 

\begin{figure*}[t!]
    \includegraphics[width=1.2\columnwidth]{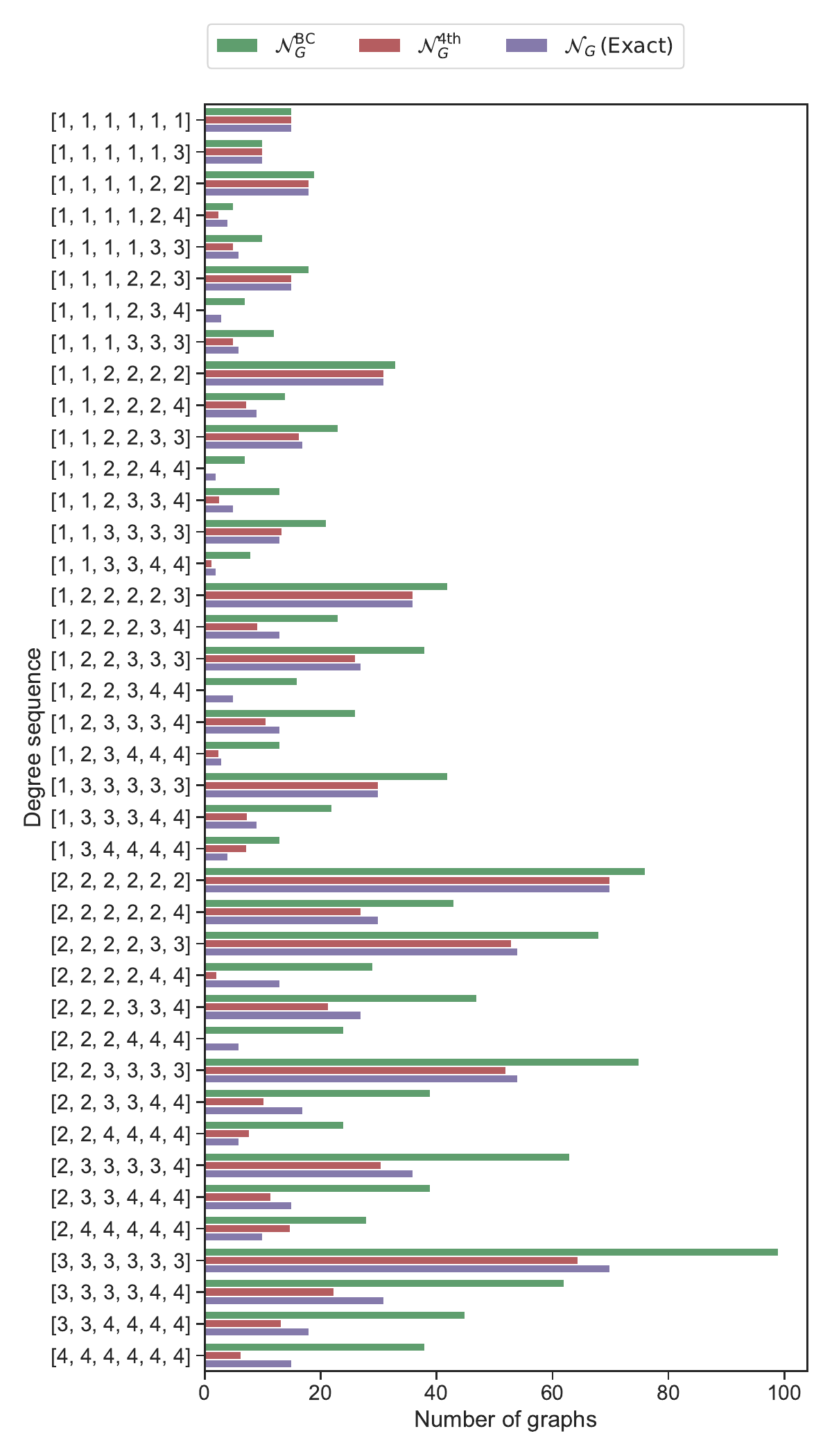}
    \caption{
    Number of graphs in a microcanonical graph ensemble with a prescribed degree sequence. 
    The graphs with $N=6$ are considered and the degree sequences are those with $\mathcal{N}_{G}(\ket{d})>0$. 
    The estimates $\mathcal{N}^{\mathrm{4th}}_{G}(\ket{d})$ and $\mathcal{N}^{\mathrm{BC}}_{G}(\ket{d})$ are rounded to integers. 
    }
    \label{fig:comparison}
\end{figure*}

Next, the dependence of the degree distribution is evaluated on the estimates $\mathcal{N}^{\mathrm{4th}}_{G}(\ket{d})$ and $\mathcal{N}^{\mathrm{BC}}_{G}(\ket{d})$ for $N=6$. 
Herein, all possible graphic degree sequences of six vertices are considered, that is, the sum of degrees is even and the Erd\H{o}s--Gallai condition \cite{ErdosGallai1960} is satisfied. 
Figure~\ref{fig:comparison} shows the results with such degree sequences. 
Without the loss of generality, each degree sequence can be assumed to have a non-descending order because $\mathcal{N}_{G}(\ket{d})$ is invariant under the permutation of elements within a degree sequence. 
Again, the estimate $\mathcal{N}^{\mathrm{4th}}_{G}(\ket{d})$ is often closer to the exact value $\mathcal{N}_{G}(\ket{d})$ than $\mathcal{N}^{\mathrm{BC}}_{G}(\ket{d})$. 
The results also indicate that, although $\mathcal{N}^{\mathrm{4th}}_{G}(\ket{d})$ tends to underestimate $\mathcal{N}_{G}(\ket{d})$, it is not a lower bound. 
By contrast, $\mathcal{N}^{\mathrm{BC}}_{G}(\ket{d})$ overestimates $\mathcal{N}_{G}(\ket{d})$. 

\section{Discussion}\label{sec:Discussion}
To obtain the entropy of the finite-graph ensembles, we estimated the total number of instances $\mathcal{N}_{G}(\ket{d})$ by expanding the generating function. 
In statistical physics, although generating function techniques are common, a large-size (or a thermo-dynamic) limit is usually considered, and the saddle-point approximation is often applied. 
By contrast, we established a mean-field approach to evaluate finite-size graphs using the {\CH} polynomials. 

While many of discrete system problems are combinatorial, the combinatorial aspect is often more pronounced when the system is small. 
Therefore, the method used in this study, which allowed us to observe the connection to combinatorial evaluations, is particularly important for ensembles of small systems. 
Even when the number of vertices $N$ is small, $\mathcal{N}_{G}(\ket{d})$ can be considerably large, and its accurate estimation can be computationally demanding. 

Although we analyzed up to the incomplete fourth-order expansion of the generating function, it is not clear whether even finer expansions still yield concise solutions that can be interpreted in a combinatorial manner. 
Furthermore, it would be interesting to determine whether a mean-field estimate can be sufficiently sophisticated to distinguish a non-graphic degree sequence as $\mathcal{N}_{G}(\ket{d})=0$. 

Quantitatively estimating the accuracy of the estimation of $\mathcal{N}_{G}(\ket{d})$ is an interesting problem. 
For an ensemble of $d$-regular graphs, the multiplicative error between the {\BC} formula and $\mathcal{N}_{G}(\ket{d})$ is $O(\exp(d^{3}/N))$ \cite{McKay1985}. 
Thus, the formula is asymptotically accurate when the degree is sufficiently small. 
By contrast, our mean-field expansion does not manifestly bound $\mathcal{N}_{G}(\ket{d})$ in terms of $N$. 
Although $\mathcal{N}^{\mathrm{4th}}_{G}(\ket{d})$ is relatively accurate, it was confirmed that it is not a bound of $\mathcal{N}_{G}(\ket{d})$. 

This study focused on microcanonical graph ensembles with relatively simple constraints. 
We believe that an analogous analysis can be performed for graph ensembles that allow multi-edges and self-loops. 
Furthermore, we expect that the mean-field approach considered in this study can be applied to various problems involving finite discrete systems.

\begin{acknowledgments}
The author thanks Oleg Evnin and Sergei Nechaev for fruitful discussions.
This work was supported by JST ACT-X Grant No. JPMJAX21A8 and JSPS KAKENHI No. 22H00827.
\end{acknowledgments}

%


\end{document}